\documentclass[twocolumn,letter]{jpsj2} 
\def\lsim{\mathstrut_{\displaystyle \sim}^{\displaystyle <}}
\def\gsim{\mathstrut_{\displaystyle \sim}^{\displaystyle >}}

\title{
Kink Structure in the Quasiparticle Band of Doped Hubbard Systems
}

\author{
Yoshiro \textsc{Kakehashi}$^{1}$\thanks{E-mail address:
yok@sci.u-ryukyu.ac.jp} and Peter \textsc{Fulde}$^{2}$\thanks{E-mail
address: fulde@mpipks-dresden.mpg.de}
}

\inst{
$^{1}$Department of Physics and Earth Sciences,
Faculty of Science, University of Ryukyus,
1 Senbaru, Nishihara, Okinawa 903-0213, Japan \\
$^{2}$Max-Planck-Institut f\"ur Physik komplexer Systeme,
N\"othnitzer Str. 38, D-01187 Dresden, Germany
}

\abst{
By making use of the self-consistent projection operator method with 
high-momentum and high-energy resolutions, we find a kink structure in the
quasiparticle excitation spectrum of the two-dimensional Hubbard model 
in the underdoped regime. The kink is caused by a mixing
between the quasiparticle state and excitations with
short-range antiferromagnetic order.  We suggest that this
might be the origin of the strong concentration dependence of the 'kink' found
in La${}_{2-x}$Sr${}_{x}$CuO${}_{4}$ ($x=0.03-0.07$).
}

\kword{kink, quasiparticle excitations, Hubbard model, ARPES, cuprate, 
LSCO}

\begin{document}
\maketitle

Recent high-resolution photoemission experiments show that there 
is a well-defined '{\it kink}' in the quasiparticle band dispersion of
high-T${}_{\rm c}$ cuprates, whose energy scale is 
$\omega_{\rm kink} = 60-70$ meV in both the normal and superconducting
states \cite{bodanov,lanzara,cuk}.  
The kink along the nodal direction was found to have a
universal feature \cite{lanzara}, i.e., 
the Fermi velocity $v_{\rm F}$ in the low-energy
region ($|\omega| < \omega_{\rm kink}$) is not sensitive to the type of
cuprates, doping concentration, and isotope substitution 
\cite{geon}, although
$v_{\rm F}$ in the high-energy regime ($|\omega| > \omega_{\rm kink}$)
strongly depends on the latter.
Various theoretical explanations for the kink have been attempted from 
two different points of view.  One type of theory relies on the coupling 
of an electronic quasiparticle to the spin fluctuation resonance 
mode observed in inelastic
neutron scattering experiments \cite{eschrig,johnson}.  Another relies on
a coupling to phonons, particularly to the longitudinal optical phonon mode
found in neutron experiments \cite{lanzara,cuk}.  
The latter approach raised again the fundamental question on 
the mechanism underlying  high-T${}_{\rm c}$ superconductivity 
in cuprates, i.e., 
electron- or phonon-mediated.  Although the reliability of these
theories is under debate and the improvements of these theories are in
progress \cite{schach,ishihara}, 
it has not yet been seriously studied whether or not the kink
in the quasiparticle state can be solely of electronic origin.
We deal with this problem in the present letter and report the
appearance of a kink due to long-range electron correlations in the
two-dimensional (2D) Hubbard model for small doping concentrations.

The difficulty in the present problem is that the perturbation approach 
is not applicable to cuprates because of strong electron 
correlations. Therefore, advanced theories such as the Lanczos method
\cite{dagotto94}, the quantum Monte-Carlo (QMC) method
\cite{bulut94,preuss94,grober00}, and dynamical cluster approximation (DCA)
\cite{jarrell01,maier02}, have been applied to the cuprate system.  They
clarified the global structure of the single-particle excitation spectrum in
the 2D Hubbard model. A detailed structure of the low-energy 
excitations at low temperatures, however, has not been derived because of 
the limited resolutions in both momentum and energy and the limited 
range of intersite electron correlations inherent in 
the cluster approaches.  In particular, a high resolution is 
indispensable for theoretical investigations of the kink.
In order to overcome these difficulties, we have recently developed the
self-consistent projection operator method (SCPM) on the basis of the
retarded Green function \cite{kake04-1,kake04-3}. The method is suitable for
the present purpose because it allows for calculating self-energy with 
high-momentum and high-energy resolutions, and self-consistently taking 
into account the
long-range intersite correlations by using an incremental cluster expansion
and an effective medium.  The results presented here were obtained by 
this method.

We apply the 2D Hubbard model on the square lattice using the 
nearest-neighbor electron hopping parameter $t$ and intra-atomic 
Coulomb interaction parameter $U$.  The single-particle excitation 
energy spectrum is obtained from the retarded Green function 
\begin{equation}
G_{\mbox{\boldmath$k$}}(z) = \frac{1}{z - \epsilon_{\mbox{\boldmath$k$}} 
- \Lambda_{\mbox{\boldmath$k$}}(z)} \ .
\label{gk}
\end{equation}
Here, $z=\omega + i \delta$, where $\delta$ is a positive infinitesimal
number, and $\epsilon_{\mbox{\boldmath$k$}}$ is the Hartree-Fock
one-electron energy dispersion measured from Fermi energy.  
In the SCPM,
the momentum-dependent self-energy $\Lambda_{\mbox{\boldmath$k$}}(z)$ 
is calculated from the Fourier transform of nonlocal memory functions 
$M_{ij}$ as
\begin{equation}
\Lambda_{\mbox{\boldmath$k$}}(z)=U^{2}\sum_j M_{j0}(z) 
\exp (i \mbox{\boldmath$k$} \cdot \mbox{\boldmath$R$}_j) \ .
\label{lk}
\end{equation}
Note that $\mbox{\boldmath$R$}_j$ is the position vector of site $j$. 
High-momentum and high-energy resolutions are achieved by taking into account
the off-diagonal terms $M_{ij}(z)$ up to infinity.

We calculate $M_{ij}(z)$ by means of an incremental cluster expansion in an
effective medium with a coherent potential $\tilde{\Sigma}(z)$. Within the
two-site approximation, the $M_{ij}(z)$ are given by 
$M_{ii}(z) =  M^{(i)}_{ii}(z) +
\sum_{l \neq i} \left( M^{(il)}_{ii}(z)-M^{(i)}_{ii}(z)\right)$ and
$M_{i \neq j}(z) = M^{(ij)}_{i \neq j}(z)$. $M^{(i)}_{ii}(z)$ and $M^{(ij)}_{i
  \neq j}(z)$ are the matrix elements of the cluster memory matrices defined by
$M^{(c)}_{lm}(z) = \big[ \hat{\mbox{\boldmath$M$}}^{(c)} \big( 1 -
  {\mbox{\boldmath$L$}}^{(c)} \cdot \hat{\mbox{\boldmath$M$}}^{(c)} \big)^{-1}
  \big]_{lm}$ $(c=i, ij)$.  Here, ${\mbox{\boldmath$L$}}^{(c)}(z)$ is a
1$\times$1 (for $c=i$) or (2$\times$2) ($c=(ij)$) cluster Liouvillean, whose
diagonal matrix elements are given by $L^{(i)}(z)=U(1-2\langle
n_{i-\sigma}\rangle)/[\langle n_{i-\sigma} \rangle (1 - \langle
  n_{i-\sigma}\rangle)]$ using an average electron number  $\langle n_{i
  \sigma} \rangle$ with spin $\sigma$ on site $i$.
The screened memory matrix element $\hat{M}^{(c)}_{ij}(z)$ is obtained
using renormalized perturbation theory \cite{kake04-1} as
\begin{equation}
\hat{M}^{(c)}_{ij}(z) = A_{ij} \int \frac{d\epsilon d\epsilon^{\prime} 
d\epsilon^{\prime\prime} \tilde{\rho}^{(c)}_{ij}(\epsilon)
\tilde{\rho}^{(c)}_{ij}(\epsilon^{\prime})
\tilde{\rho}^{(c)}_{ji}(\epsilon^{\prime\prime}) \chi(\epsilon,
\epsilon^{\prime}, \epsilon^{\prime\prime})}
{z  - \epsilon - \epsilon^{\prime} + \epsilon^{\prime\prime}}~, 
\label{lrpt}
\end{equation}
where 
$A_{ii}=[\langle n_{i-\sigma} \rangle (1 - \langle
n_{i-\sigma}\rangle)]/[\langle n_{i-\sigma} \rangle_c (1 - \langle
n_{i-\sigma}\rangle_c)]$ and $A_{i \neq j}=1$. Furthermore,
$\langle n_{i \sigma}
\rangle_c=\int d\epsilon \tilde{\rho}^{(c)}_{ii}(\epsilon) f(\epsilon)$ 
with the Fermi distribution function
$f(\epsilon)$. Moreover, 
$\chi(\epsilon, \epsilon',\epsilon'')=f(-\epsilon) f(-\epsilon') 
f(\epsilon'') + f(\epsilon) f(\epsilon')
f(-\epsilon'')$.  
The matrix $\tilde{\rho}^{(c)}_{ij}(\epsilon)$ is the density of states 
of a system with an empty site $i$ (or sites $i$ and $j$) embedded in 
a medium $\tilde{\Sigma}(z)$. The latter is determined self-consistently 
by means of the coherent potential approximation (CPA) equation 
\cite{kake05}.
We note that the self-energy obtained using eq. (\ref{lrpt})
yields the second-order perturbation theory in the limit of a small
$U$ and reduces to the exact result in the limit of a large $U$.
In the intermediate coupling regime, we find a quantitative agreement
of quasi-particle bands between the SCPM and QMC method as will be shown
below.
\begin{figure}[tb]
 \begin{center}
   \includegraphics[width=8.5cm]{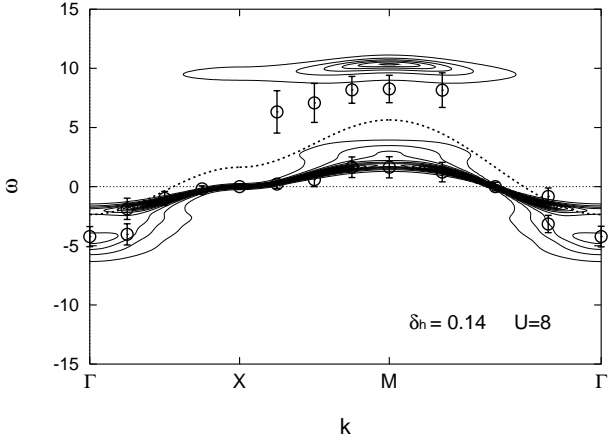}
 \end{center}
\caption{
Excitation spectra along high symmetry line calculated at hole 
concentration $\delta_{\rm h}=0.14$, $U=8$, and
 $T=0$. $\Gamma=(0,0)$, X = $(0,\pi)$, and M = $(\pi,\pi)$ in units of
 the lattice constant.  The energy unit is chosen
 so that the nearest-neighbor transfer integral is one.  Open circles
 with error bars are the results obtained by the QMC method
 \cite{grober00} at $T=0.33$.  The dashed
 curve shows the Hartree-Fock contribution $\epsilon_{k}$.
}
\label{cntr86}
\end{figure}

In the numerical calculations, we adopted the two-site approximation and
took into account intersite correlations up to the 50th nearest
neighbor site.  Moreover, we assumed excitations from the
paramagnetic ground state ({\it i.e.}, the normal state).  
Figure 1 shows the momentum-dependent excitation spectrum of a doped Hubbard
system for an intermediate Coulomb interaction strength $U=8|t|$ at $T=0$ and
a hole concentration $\delta_{\rm h}=0.14$. The excitations consist of an upper
Hubbard band around the M $(\pi,\pi)$ point, incoherent excitations at the
$\Gamma$ point with an energy $\omega \sim 4|t|$ as traces of the lower Hubbard
band, and the quasi-particle band near the Fermi level.  We emphasize that the
quasi-particle band agrees well with that obtained by the QMC method
 \cite{grober00} for the same value of $U$.  We do not find, 
however, any kink in the quasi-particle dispersion in the optimal 
doping regime, although the resolution of our results is much improved 
as compared with that obtained by the QMC method.
\begin{figure}[tb]
 \begin{center}
   \includegraphics[width=8.5cm]{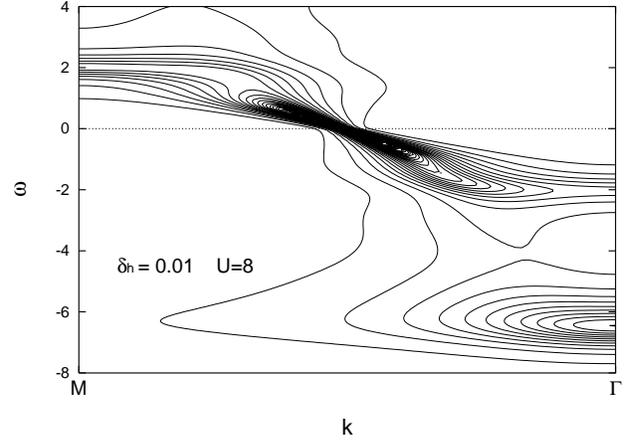}
 \end{center}
\caption{
Contour map of excitation spectrum along nodal direction at 
$\delta_{\rm h}=0.01$
}
\label{cntr99}
\end{figure}
\begin{figure}[tb]
 \begin{center}
   \includegraphics[width=8.5cm]{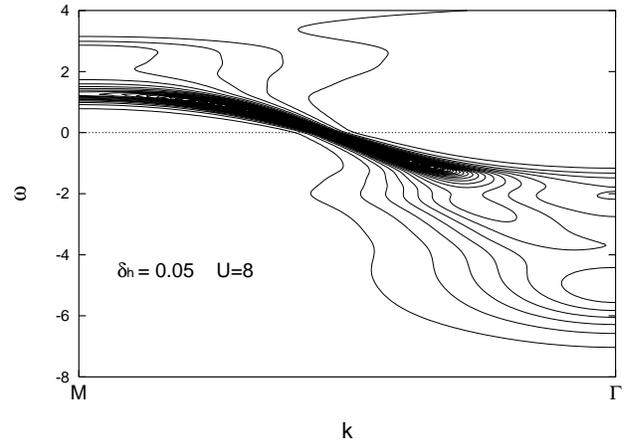}
 \end{center}
\caption{
Contour map of excitation spectrum along nodal direction at 
$\delta_{\rm h}=0.05$
}
\label{cntr95}
\end{figure}

We have examined the details of momentum dependent excitations along the nodal
direction ($(0,0)-(\pi,\pi)$) from the underdoped to the overdoped regime.
Figure 2 shows the results for a doping concentration of $\delta_{\rm h} =
0.01$.  We find here a kink at $|\mbox{\boldmath$k$}|=0.5\pi$ and $\omega_{\rm
  kink}=-0.8|t|$. The kink becomes weaker with increasing
doping concentration and the velocity ratio of the high-energy part to
the low-energy one ($v^{\prime}_{\rm F}/v_{\rm F}$) at the kink 
position becomes smaller ({\it e.g.}, 
$v^{\prime}_{\rm F}/v_{\rm F}=1.8$ for $\delta_{\rm h}=0.01$, 
and $v^{\prime}_{\rm F}/v_{\rm F}=1.5$ for $\delta_{\rm h}=0.02$), 
while the position hardly changes with $\delta_{\rm h}$.  
As shown in Fig. 3, the kink disappears for $\delta_{\rm h} = 0.05$. 
This occurs together with a collapse of the lower Hubbard band.

It should be noted that in Fig. 2 the flat band exists at $\omega=\pm 2.0|t|$ 
because of the excitations due to short-range magnetic order ({\it i.e.}, a
precursor of the gap formation due to antiferromagnetic correlations)
\cite{preuss94,jarrell01}.  The kink at $\omega_{\rm kink}=-0.8|t|$ is caused
by a mixing between the quasiparticle state and magnetic excitations.
Recently, we have reported that a marginal Fermi liquid \cite{varma89} 
(MFL)-like behavior is found in the underdoped region ($\delta_{\rm h} \ \lsim
\ 0.03$) away from half-filling because of a pinning of Fermi energy to the
van Hove anomaly due to a transfer of spectral weight from the lower Hubbard
band to the upper one \cite{kake05}.  
Antiferromagnetic correlations should be enhanced in this region because of
nesting. The kink behavior appears in this region $\delta_{\rm h} \ \lsim \
0.03$. On the other hand, the lower Hubbard band collapses when $\delta_{\rm h}
\ \gsim \ 0.03$. The MFL-like state with antiferromagnetic correlations changes
to a normal Fermi liquid state and the kink disappears, as noted in
Fig. 3. The calculated Fermi velocity along the $\Gamma$-M line is
presented in Fig. 4.  The velocity shows a weak concentration
dependence, and the renormalization factor of the Fermi velocity is about
1.8 from the underdoped region to the overdoped one.
\begin{figure}[tb]
 \begin{center}
   \includegraphics[width=8.5cm]{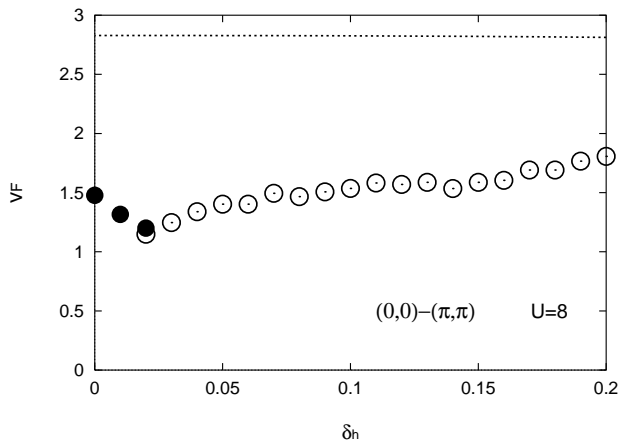}
 \end{center}
\caption{
Fermi velocity along nodal direction as function of hole 
concentration.  Closed circles denote the velocity in the marginal Fermi
 liquid state, while open circles indicate the velocity in the normal
 Fermi liquid state.  The dashed line is the result for a noninteracting
 system. 
}
\label{vf}
\end{figure}

We have also examined in detail excitations along $\Gamma$-X-M.
Because of the van Hove singularity, the quasiparticle band around the X
point is quite flat.  We do not find a kink behavior in this flat
band region.  The mixing between the quasiparticle band and the band of the
magnetic excitations takes place away from the linear dispersion regime near
the Fermi level, i.e., $(|\mbox{\boldmath$k$}|,\omega)=(0.5\pi,-1.2)$ on the
$\Gamma$-X line. Because both bands are flat, it is not clear whether this
region contains a kink.   

The present model is too simple for attempting a comparison with 
the experimental data.  
Nevertheless, it is plausible that the kink is enhanced with
decreasing doping concentration because of the development of short-range
antiferromagnetic order.  We speculate that the strong concentration dependence
of the kink in La${}_{2-x}$Sr${}_{x}$CuO${}_{4}$ 
(LSCO) \cite{cuk,zhou} may be caused by the present mechanism.  In fact, we
obtain for the characteristic kink energy $\omega_{\rm kink}=70$ meV when we
choose the transfer integral so that the calculated Fermi velocity along the 
nodal direction agrees with the observed one (1.8 eV$\cdot$\AA).
The value for $\omega_{\rm kink}$ agrees well with the experimental one (60-70
meV) \cite{zhou}. 

In summary, we have investigated the quasiparticle band in the doped 2D
Hubbard model on the basis of the SCPM with high-momentum and
high-energy resolutions.  We find a kink along the nodal direction 
$(0,0)-(\pi,\pi)$ in the range of doping concentrations 
$0 < \delta_{\rm h} \ \lsim \ 0.03$ where the MFL behavior persists.  
It is caused by a mixing between the quasiparticle excitations and magnetic
excitations with short-range antiferromagnetic order.  The kink decays rapidly
with the decrease in antiferromagnetic correlations. We speculate that the
kink in the underdoped regime of LSCO may be due to the present mechanism.

\section*{Acknowledgements}
The authors would like to thank Drs. J. Fink and O. Gunnarsson for 
valuable discussions.


\begin{thebibliography}{99} 
%
\bibitem{bodanov}
P. V. Bodanov, A. Lanzara, S. A. Keller, X. J. Zhou, E. D. Lu, W. J. Zheng, 
G. Gu, J. -I. Shimoyama, K. Kishio, H. Ikeda, R. Yoshizaki,
Z. Hussain and Z. X. Shen: Phys. Rev. Lett. {\bf 85} (2000) 2581.
%
%
\bibitem{lanzara}
A. Lanzara, P. V. Bogdanov, X. J. Zhou, S. A. Keller, D. L. Feng, E. D. Lu,
T. Yoshida, H. Eisaki, A. Fujimori, K. Kishio, J. -I. Shimoyama,
T. Noda, S. Uchida, Z. Hussain and Z. -X. Shen: Nature {\bf 412} (2001) 510.
%
%
\bibitem{cuk}
T. Cuk, D.H. Lu, X.J. Zhou, Z.-X. Shen, T.P. Devereax and N. Nagaosa: 
Phys. Stat. Sol. (b) {\bf 242} (2005) 11.
%
%
\bibitem{geon}
G.-H. Gweon, T. Sasagawa, S.Y. Zhou, J. Graf, H. Takagi, D.-H. Lee and
A. Lanzara: Nature {\bf 430} (2004) 187.
%
%
\bibitem{eschrig}
M. Eschrig and M. R. Norman: Phys. Rev. Lett. {\bf 85} (2000) 3261;
Phys. Rev. Lett. {\bf 89} (2002) 277005.
%
%
\bibitem{johnson}
P. D. Johnson, T. Valla, A. V. Fedorov, Z. Yusof, B. O. Wells, Q. Li,
A. R. Moodenbaugh, G. D. Gu, N. Koshizuka, C. Kendziora, Sha Jian
and D. G. Hinks: Phys. Rev. Lett. {\bf 87} (2001) 177007.
%
%
\bibitem{schach}
E. Schachinger, J. J. Tu and J.P. Carbotte: cond-mat/0304029.
%
%
\bibitem{ishihara}
S. Ishihara and N. Nagaosa: Phys. Rev. B {\bf 69} (2004) 144520.
%
%
\bibitem{dagotto94}
E. Dagotto: Rev. Mod. Phys. {\bf 66} (1994) 763.
%
%
\bibitem{bulut94}
N. Bulut, D. J. Scalapino and S. R. White: Phys. Rev. Lett. {\bf 73}
(1994) 748; {\bf 72} (1994) 705; Phys. Rev. {\bf 50} (1994) 7215.
%
\bibitem{preuss94}
R. Preuss, W. Hanke and W. von der Linden: Phys. Rev. Lett. {\bf 75}
(1994) 1344.
%
%
\bibitem{grober00}
C. Gr\"ober, R. Eder and W. Hanke: Phys. Rev. B{\bf 62} (2000) 4336.
%
%
\bibitem{jarrell01}
M. Jarrell, Th. Maier, C. Huscroft and S. Moukouri: Phys. Rev. B {\bf
 64} (2001) 195130.
%
%
\bibitem{maier02}
Th. A. Maier, Th. Pruschke and M. Jarrell: Phys. Rev. B{\bf 66}
 (2002) 075102. 
%
%
\bibitem{kake04-1}
Y. Kakehashi and P. Fulde: Phys. Rev. B{\bf 70} (2004) 195102.
%
%
\bibitem{kake04-3}
Y. Kakehashi: Adv. Phys. {\bf 53} (2004) 497.
%
%
\bibitem{varma89}
C. M. Varma, P. B. Littlewood, S. Schmitt-Rink, E. Abrahams and
 A. E. Ruckenstein: Phys. Rev. Lett. {\bf 63} (1989) 1996.
%
%
\bibitem{kake05}
Y. Kakehashi and P. Fulde: Phys. Rev. Lett. {\bf 94} (2005) 156401.
%
%
\bibitem{zhou}
X. J. Zhou, J. Shi, T. Yoshida, T. Cuk, W. L. Yang, V. Brouet,
J. Nakamura, N. Mannella, S. Komiya, Y. Ando, F. Zhou, W. X. Ti,
J. W. Xiong, Z. X. Zhao, T. Sasagawa, T. Kakeshita, E. Eisaki,
S. Uchida, A. Fujimori, Z. Zhang, E. W. Plummer, R. B. Laughlin,
Z. Hussain and Z. -X. Shen: cond-mat/0405130.
%
%
\end{thebibliography}
\end{document}